\documentclass[preprint,12pt]{elsarticle}

\usepackage{amssymb}
\usepackage{float}	
\usepackage{amsmath}	
\usepackage{color}
\usepackage{graphicx}
\usepackage{amsfonts}
\usepackage{epstopdf}
\usepackage[utf8]{inputenc}
\usepackage[T1]{fontenc}

\begin{document}

\begin{frontmatter}

\title{Galilean-transformed solitons and supercontinuum generation in dispersive media}


\address[label1]{Centre for Wind, Waves and Water, School of Civil Engineering, The University of Sydney, Sydney, Australia}
\address[label2]{LHEEA, \'{E}cole Centrale Nantes, UMR CNRS No. 6598, Nantes, France}
\address[label3]{Dynamics Group, Hamburg University of Technology, Hamburg, Germany}
\address[label4]{Department of Mechanical Engineering, Imperial College London, London, United Kingdom}
\address[label5]{Institut FEMTO-ST, UMR 6174 CNRS-Universit\'{e} de Franche-Comt\'{e}, Besan\c{c}on, France}
\address[label6]{Disaster Prevention Research Institute, Kyoto University, Kyoto, Japan}
\address[label7]{Hakubi Center for Advanced Research, Kyoto University, Kyoto, Japan}

\author[label1]{Y. He}
\ead{yuchen.he@sydney.edu.au}
\author[label2]{G. Ducrozet} 
\author[label3,label4]{N. Hoffmann}
\author[label5]{J. M. Dudley}
\author[label1,label6,label7]{A. Chabchoub}

\begin{abstract}
The Galilean transformation is a universal operation connecting the coordinates of a dynamical system, which move relative to each other with a constant speed. In the context of exact solutions of the universal nonlinear Schrödinger equation (NLSE), inducing a Galilean velocity (GV) to the pulse involves a frequency shift to satisfy the symmetry of the wave equation. As such, the Galilean transformation has been deemed to be not applicable to wave groups in nonlinear dispersive media. In this paper, we demonstrate that in a wave tank generated Galilean transformed envelope and Peregrine solitons show clear variations from their respective pure dynamics on the water surface. The type of deviations depends on the sign of the GV and can be captured by the modified NLSE or the Euler equations. Moreover, we show that positive Galilean-translated envelope soliton pulses exhibit self-modulation. While designated GS and wave steepness values expedite multi-soliton dynamics, the strong focusing of such higher-order coherent waves inevitably lead to the generation of supercontinua as a result of soliton fission. We anticipate that kindred experimental and numerical studies  might be implemented in other dispersive wave guides governed by nonlinearity. 

\end{abstract}

\end{frontmatter}

\section{Introduction}
The nonlinear Schrödinger equation (NLSE) is a comprehensive framework which describes the dynamics of wave pulses beyond surface gravity water waves \cite{yuen_nonlinear_1982,osborne2010nonlinear}. For instance, in Kerr media, plasma, and Bose-Einstein condensates \cite{remoissenet2013waves,dudley2014instabilities,carr2004spontaneous}. Since the proof of NLSE-integrability \cite{shabat_exact_1972}, several key wave envelope solutions have been derived and discussed within the context of modulation instability, see  \cite{akhmediev1997solitons,kivshar2003optical,malomed2006soliton,ablowitz_nonlinear_2011,he2013generating,onorato2013rogue,dudley_rogue_2019}. Besides the time-reversal symmetry \cite{chabchoub2014time}, which has been proven to be useful for applications \cite{ducrozet2020experimental}, another invariant operation is the  Galilean transformation (GT) \cite{akhmediev1997solitons,dysthe_note_1999,widjaja2021absence}. That said, introducing a Galilean velocity (GV) to a NLSE pulse is colligated to a carrier frequency-shift to satisfy symmetry \cite{sulem2007nonlinear}. As such, the GT has been legitimately considered not to be applicable and relevant for dispersive physical systems \cite{akhmediev1997solitons,christov2001energy,duran2013galilean}. 

In this work, we report experimental observations of Galilean NLSE solitons for a wide range of GVs without an external flow forcing \cite{rozenman2020observation}. The experimental data show considerable deviations from hydrodynamic NLSE predictions, nevertheless, a very good agreement with the modified NLSE (MNLSE) \cite{dysthe_note_1979} and the numerically-solved Euler equations using the higher-order spectral method (HOSM) \cite{dommermuth1987high,ducrozet2012modified}. It is shown that the type of deviations in wave envelope depends on the sign of GV. For the case of the envelope soliton, the pulse undergoes a strong broadening when GV is negative, whereas a strong self-focusing of wave envelope is observed when the GV is positive \cite{cousins2015unsteady}. Furthermore, we show that for predetermined carrier wave steepness parameters, there are exact GV values for which a Galilean-transformed envelope soliton corresponds to an exact multi-soliton solution \cite{satsuma1974b}. In fact, the higher the value of GV, the higher the order of the multi-soliton. As a result, supercontinuum generation is an unavoidable process, which follows as a result of substantial wave focusing and subsequent soliton fission events \cite{dudley2006supercontinuum,dudley2013supercontinuum,chabchoub2013hydrodynamic}. We experimentally confirm the hydrodynamic supercontiuum generation from Galilean envelope solitons in form of an irreversible and severe spectral broadening genesis. We also briefly discuss the influence of the GT on the Peregrine soliton dynamics, even though the limited fetch of the experimental facility does not allow the study of emergent and long-ranging complex wave patterns as a result of intrinsic and higher-order modulation instability processes \cite{erkintalo2011higher,gelash2019bound}.


\section{Constructing hydrodynamic Galilean solitons}

Our starting point is the dimensional time-like NLSE for deep-water waves. In fact, the weakly nonlinear propagation of a narrowband wave field $\psi$ with dominant wave frequency $\omega$ and wave number $k$ while propagating along the longitudinal direction $x$ follows  \cite{benney1967propagation,zakharov_stability_1968,osborne2010nonlinear}
\begin{equation}\label{eqn-dimensional NLS equation}
    i(\psi_x+\dfrac{1}{c_g}\psi_t)-\frac{k}{\omega^2}\psi_{tt}-k^3|\psi|^2\psi=0,
\end{equation}
where $\omega$ and $k$ are connected through the linear dispersion relation $\omega=\sqrt{gk}$, involving the gravitational acceleration $g$ and $t$ being the time coordinate while $c_g=\dfrac{\omega}{2k}$ denotes the group velocity in deep-water. The simplicity of this nonlinear wave framework allows for the understanding of the distinct role of dispersion and nonlinearity in the wave dynamics. Moreover, being an integrable framework \cite{shabat_exact_1972} facilitates the study of coherent structures' behavior in a variety of wave guides, regardless of being either steady or unstable \cite{onorato2013rogue,chabchoub2016hydrodynamic}. 
The dimensional form of the envelope soliton of amplitude $a$, satisfying Eq. (\ref{eqn-dimensional NLS equation}) reads 
\begin{equation}\label{original dimensional NLS soliton}
    \psi_S(x,t)=a\ {\rm sech}(-\sqrt{2}\alpha k(x-c_g t))\exp(-\frac{i\alpha^2kx}{2}),
\end{equation}
$\alpha:=ak$ being the wave steepness. Even though having a {\it simple} construction form, such localized wave groups play an important role in understanding the formation of wave coherence in the ocean \cite{slunyaev2021persistence,waseda2021directional}. Applying the GT to the sech-soliton yields to the following parametrization
\begin{equation}\label{Galileandimensional NLS soliton - with c}
\begin{split}
    \psi_{GS}(x,t)=&a\ {\rm sech}(-\sqrt{2}\alpha k(x-c_gt)-\frac{c}{2}\alpha^2kx)\exp(-\frac{i\alpha^2kx}{2})\\
    &\exp(\frac{i\sqrt{2}c\alpha k(x-c_gt)}{2}+\frac{ic^2\alpha^2kx}{8}),
 \end{split}
\end{equation}
which still satisfies the NLS Eq. (\ref{eqn-dimensional NLS equation}). We refer to \cite{akhmediev1997solitons,dysthe_note_1999} for GT details, including more convenient theoretical representations in dimensionless physical quantities. Note that the GT compromises a frequency-shift of the carrier and is different from the Tajiri-Watanabe-type construction \cite{tajiri1998breather}, which may cause experimental challenges in optical wave guides.

\begin{figure}[ht]
\centering
\includegraphics[width=1\textwidth]{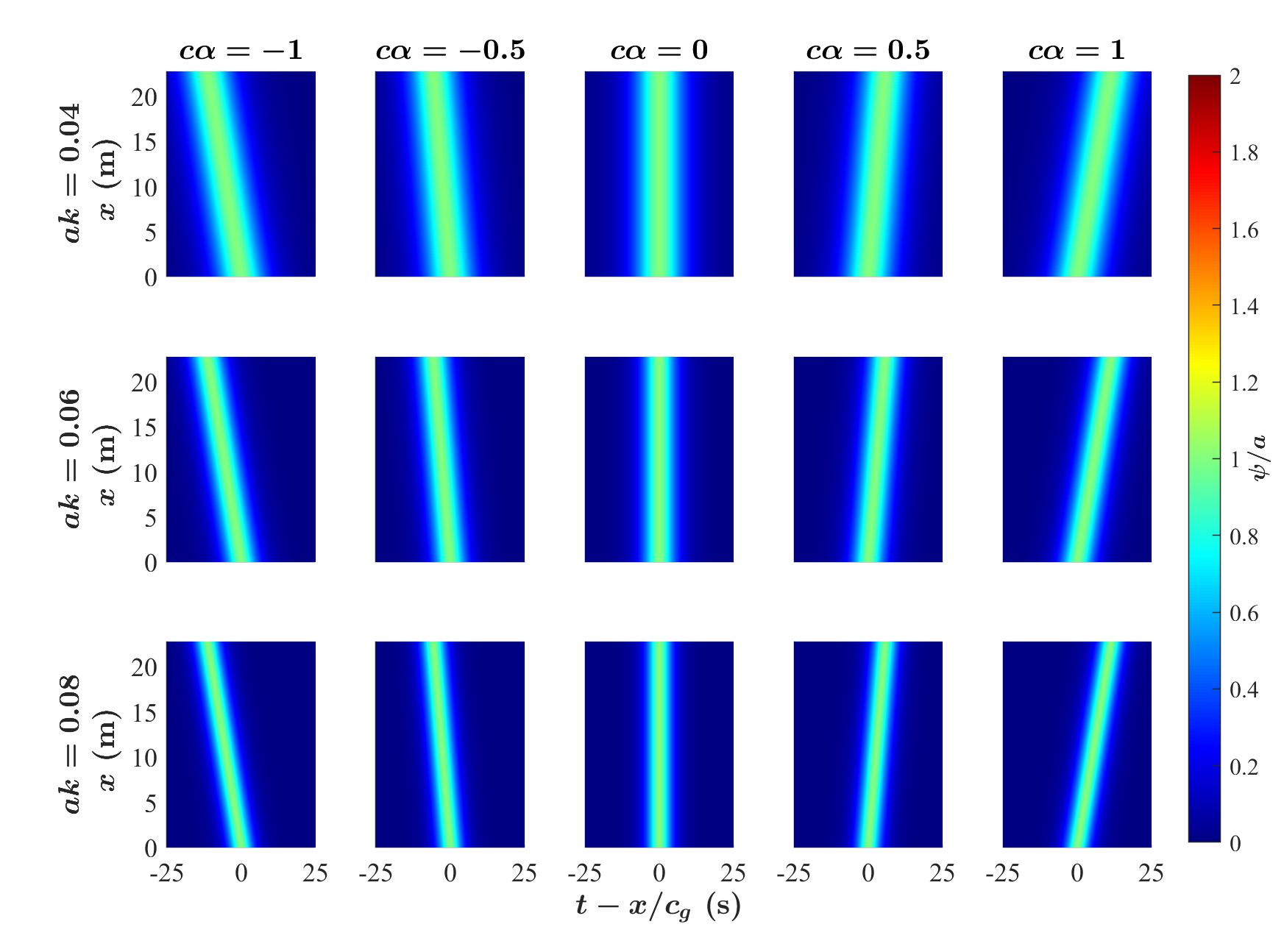}
\caption{Evolution of an NLSE envelope soliton (middle panels), slower analogues with negative GVs (left panels), and faster counterparts with positive GVs (right panels) for three wave steepness values 0.04, 0.06, and 0.08.}
\label{fig1}
\end{figure}

Examples of dimensional envelope soliton evolution  for different GVs and wave steepness values are shown in Fig. \ref{fig1}. 
We emphasize that when the GV is negative, i.e. $c<0$, the wave packet propagates slower, whereas when $c>0$ the coherent wave groups evolves faster compared to the pure and non-Galilean soliton. The Galilean-transformed hydrodynamic Peregrine soliton \cite{peregrine_water_1983} is expressed as

\begin{equation}\label{Galileandimensional NLS Peregrine breather - with c}
\begin{split}
    \psi_{GP}(x,t)=&a(-1+\frac{4(1-i\omega\alpha^2x/c_g)}{1+4(c\alpha^2kx/2+\sqrt{2}k\alpha (x-c_gt))^2+\omega^2\alpha^4x^2/c_g^2})\\
    &\exp(-i\alpha^2kx)\exp(\frac{i\sqrt{2}c\alpha k(x-c_gt)}{2}+\frac{ic^2\alpha^2kx}{8}),
 \end{split}
\end{equation}
 Both, Galilean-transformed envelope and Peregrine solitons will be tested in a water wave tank. The boundary conditions and times-series applied to the wave maker can be determined from the expression of water elevation to first-order in steepness, as defined by 
\begin{equation}
    \eta(x^\dagger,t)=\Re(\psi(x^\dagger,t)\exp(i(kx^\dagger-\omega t))), 
\end{equation}
where $x^\dagger$ can be adapted to control the location of wave focusing in the wave guide, i.e. the wave flume in our case, for the case of breathers.  

\section{Experimental build-up and arrangements}
\begin{figure}[ht]
\centering
\includegraphics[width=0.85\textwidth]{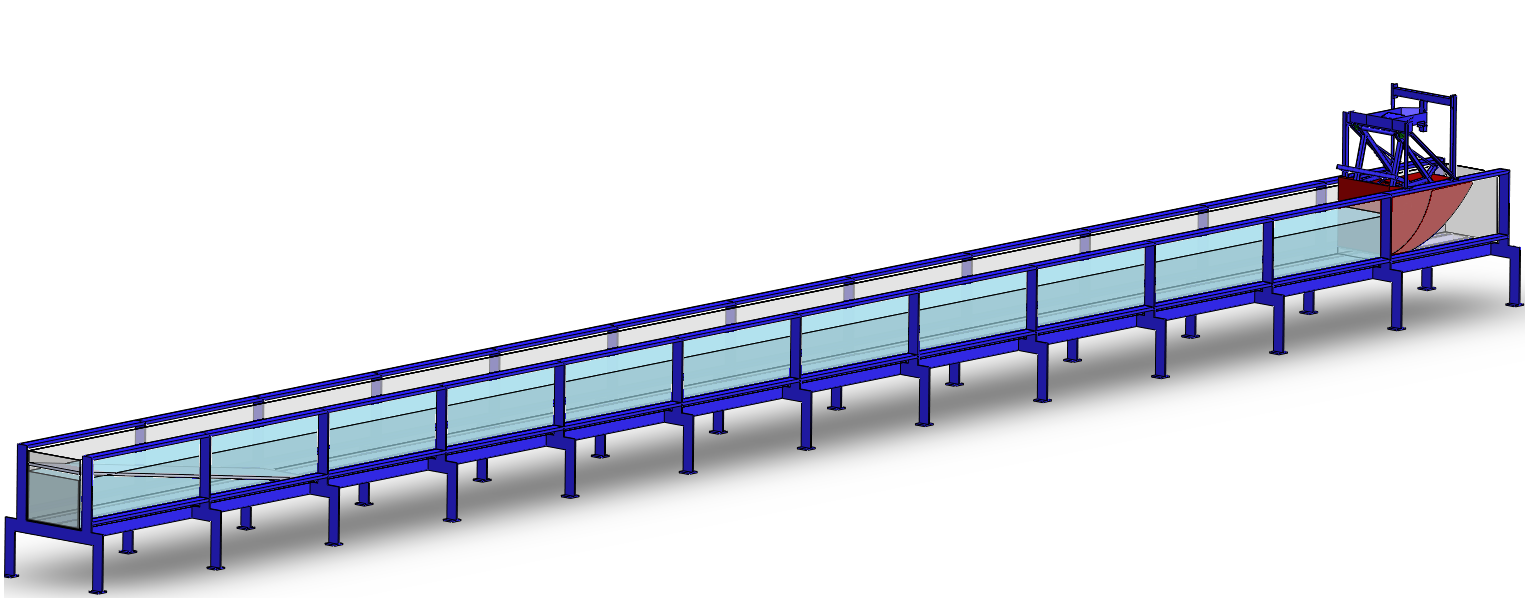}
\caption{The University of Sydney wave flume, which operates and piston wave maker.}
\label{Fig2}
\end{figure}

The experiments were carried out in the water wave flume of the University of Sydney, which is 30 meters long, 1 meter high and 1 meter wide, allowing the study of specifically controlled or irregular and random wave trains. A piston-type wave paddle located at one side of the wave tank, as shown in Fig. \ref{Fig2}, can generate waves in finite water depth as well as deep-water conditions. 
The wave paddle can generate wave time-series with a peak frequency ranging between 0.5 to 2 Hz in water depth conditions varying from 0.4 m to 0.9 m. In this study, the carrier wave frequency is fixed at $f=1.25$ Hz, which corresponds to a wavelength of $\lambda=1$ m and a wavenumber of $k=\dfrac{2\pi}{\lambda}=2\pi$. The amplitude $a$ of the carrier wave is varied to satisfy wave steepness $\alpha$ values of 0.04, 0.06 and 0.08, respectively. We also ensure that the dominant carrier frequency-shift after the GT, as defined by the GV value $c$, is within the operational range of the wave maker. The water depth is set to be $h=$ 0.7 m. Hence, $kh\gg 1$, which satisfies deep-water requirements. Eight capacitance wave gauges with a sampling frequency of 32 Hz can be re-deployed along the flume for a generated time-series to allow not only a very high temporal, but also a spatial resolution considering the repeatability of the experiments. 


\section{Experimental results}
The focus of our experimental campaign will be on the Galilean-transformed envelope solitons. The case of Peregrine breather is more complex and will be briefly discussed as well. 

\begin{figure}[ht]
\centering
\includegraphics[width=0.9\textwidth]{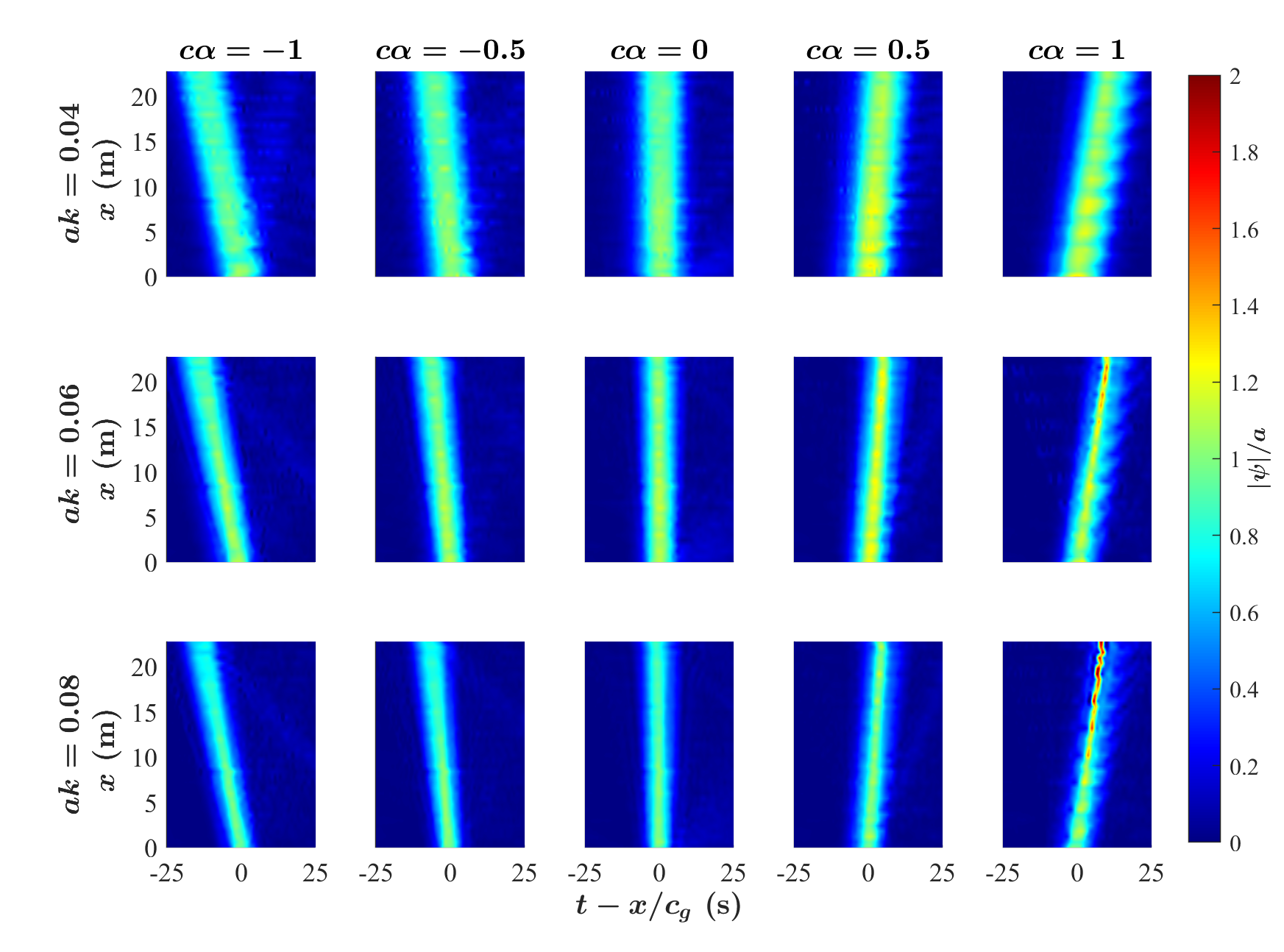}
\caption{Experimental observations of Galilean-transformed and pure envelope sech-type solitons.}
\label{Fig3}
\end{figure}

We recall that both, envelope and Peregrine solitons have been observed in a wide range of physical media \cite{yuen1975nonlinear,mollenauer1980experimental,strecker2002formation,khaykovich2002formation,kibler2010peregrine,chabchoub_rogue_2011,bailung2011observation}. In water waves envelope solitons have been confirmed to be steady even for very large steepness values, even close to the wave breaking threshold \cite{slunyaev2013simulations,toffoli2010maximum}. The carrier wave steepness does not only quantify the degree of nonlinearity of Stokes waves, but also determines the width of a wave pulse or wave packet. Three values of wave steepness have been considered $\alpha\in\{0.04,0.06,0.08\}$ as well as negative and positive values of GV. The evolution of envelope solitons subject to GT as propagating over 22 m are shown in Fig. \ref{Fig3}. 
We stress that the wave envelope has been reconstructed from the water surface data using the Hilbert transform \cite{osborne2010nonlinear}. At first view, the cases for $\alpha=0.04$ seem to show a steady evolution of the wave packet. However, since this steepness value is indeed very small, it does not allow for the nonlinear interaction to unfold over the limited fetch. Increasing the value of wave steepness reveals that differently than predicted from the NLSE framework, the Galilean-transformed solitons in a physical water wave guide are not stationary while the type of unsteadiness depends on the sign of GV. In fact, when the GV is negative, we can clearly notice that the solitons subject to GT broaden, whereas when GV is positive, the Galilean-transformed envelope solitons follow a self-compression in form of a breathing process. The latter process becomes clearly visible for the highest steepness and GV values adopted in the experiments.

To examine the physics of these deviations, we performed numerical simulations based on the modified nonlinear Schrödinger equation (MNLSE) \cite{dysthe_note_1979,lo1985numerical,trulsen2001spatial,shemer2002experimental,goullet_numerical_2011}. We recall that whereas the NLSE can be derived from the Euler equations at third-order in steepness $\mathcal{O}(\alpha^3)$ \cite{hasimoto1972nonlinear}, the MNLSE is an improvement at the next order $\mathcal{O}(\alpha^4)$, which accounts for higher-order dispersive effects \cite{dysthe_note_1979,kivshar1989dynamics,rogers2012localized}.
\begin{figure}[ht]
\centering
\includegraphics[width=0.9\textwidth]{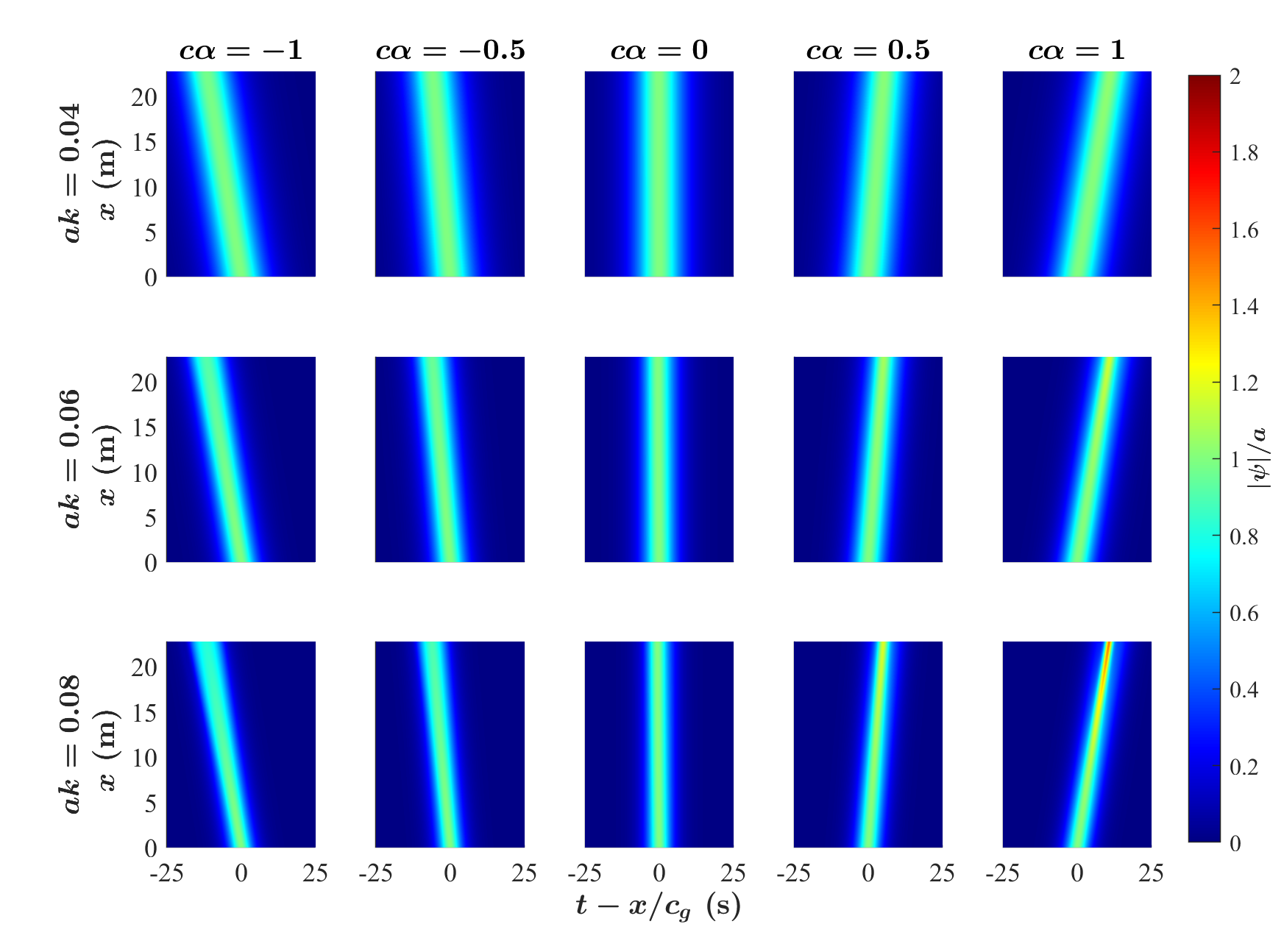}
\caption{MNLSE prediction of pure and Galilean transformed envelope solitons as generated in the wave tank and shown in Fig. \ref{Fig3}.}
\label{Fig4}
\end{figure}
The MNLSE simulation results, as revealed in Fig. \ref{Fig4}, support the evidence that unsteadiness of GT wave envelope evolution can be attributed to the failure of simplified dispersion relationship in the wave packet evolution description, similarly to the water wave propagation under the action of a uniform and steady current \cite{mei_theory_2005,kouskoulas2020linear}. 

\begin{figure}[ht]
\centering
\includegraphics[width=1\textwidth]{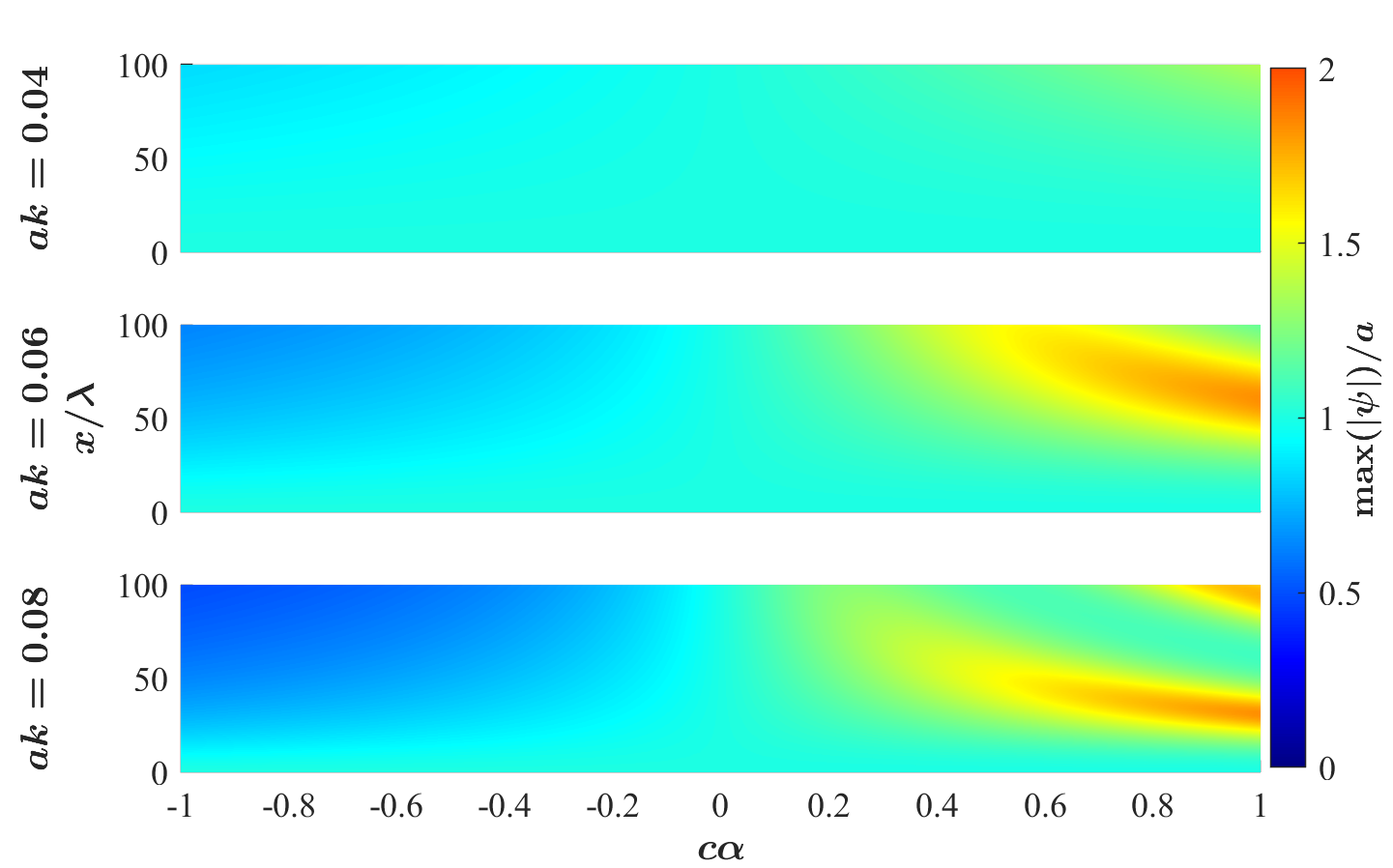}
\caption{Amplitude amplification factors of Galilean-transformed envelope solitons with respect to GV $\alpha c$ and the dimensionless evolution distance, as quantified from MNLSE simulations.}
\label{Fig5}
\end{figure}

Positive values of GVs have a significant influence on the amplitude variations of envelope solitons subject to GT, see Fig. \ref{Fig5}. 
No significant change in steadiness can be noticed for marginal $\alpha c$ values. The observed self-modulation, which has been also underlined by the MNLSE simulations, for positive GVs is indeed similar to the dynamics of multi-solitons \cite{satsuma1974b}. The connection between Galilean transformed multi-solitons and Satsuma-Yajima solutions will be further explored in the next Section 5. 

The investigation of Galilean-transformed solitons on zero-background are simpler to perform due to the absence of modulation instability \cite{bespalov1966filamentary,benjamin_disintegration_1967}. Consequently, any yet small perturbation of the background wave can provoke a drastic wave focusing \cite{suret2016single,narhi2016real,chabchoub_experiments_2017}. Moreover, the observation of such unstable patterns requires a larger fetch. This makes it challenging to be explored in the University of Sydney wave flume, which is limited to an effective propagation distance of 22 m, considering the frequency range of wave maker and the limited water depth possibilities. 

\begin{figure}[ht]
\centering
\includegraphics[width=1\textwidth]{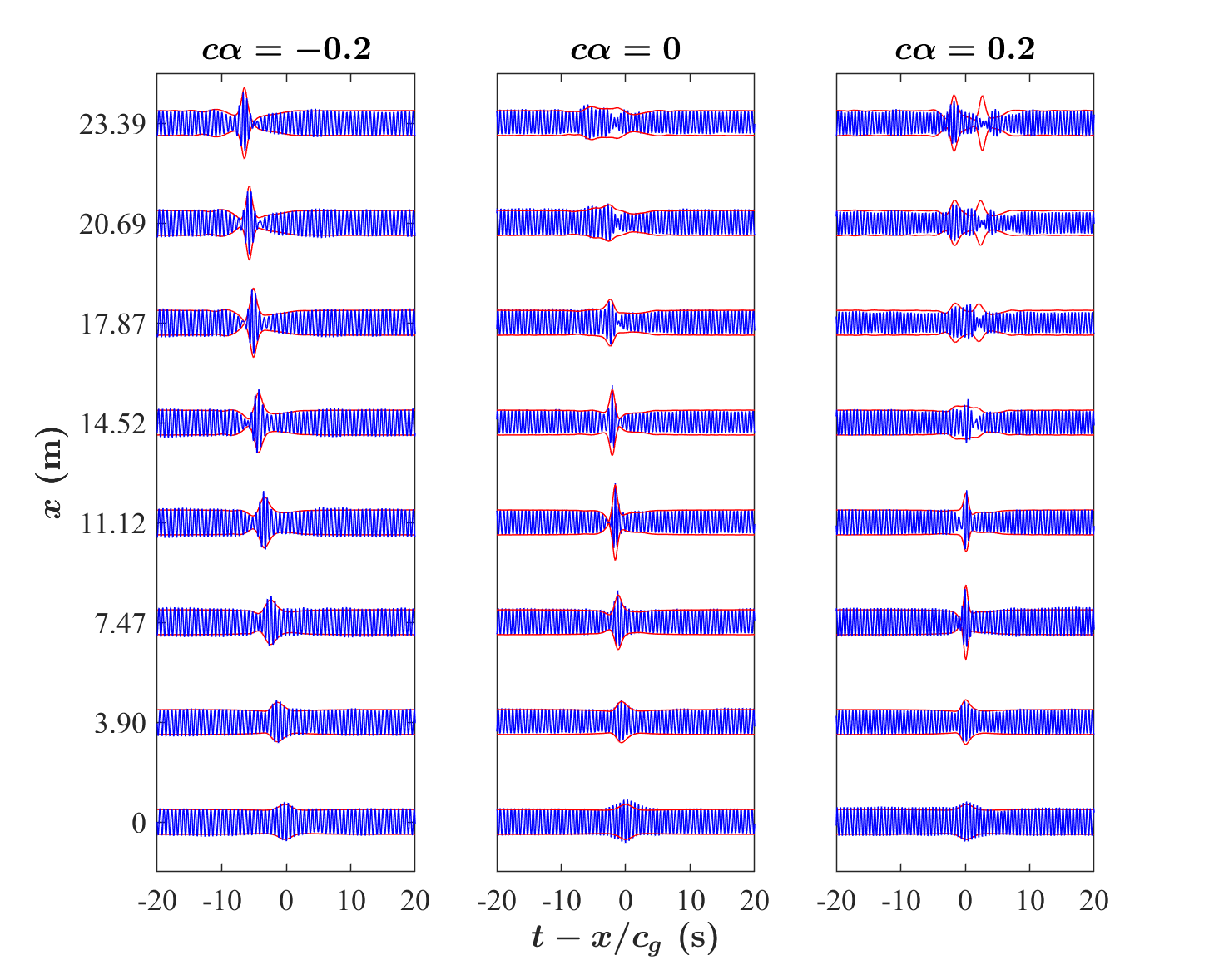}
\caption{Galilean-transformed Peregrine evolution in the wave flume (blue lines) for the carrier parameters $ak=0.1$, $a=0.01$ m and $x^\dagger=-12$ m to observe the maximal compression for the pure case in the middle of the tank. The red lines are the corresponding MNLSE simulations of wave envelope.}
\label{Fig6}
\end{figure}

\begin{figure}[ht]
    \centering
    \includegraphics[width=1\textwidth]{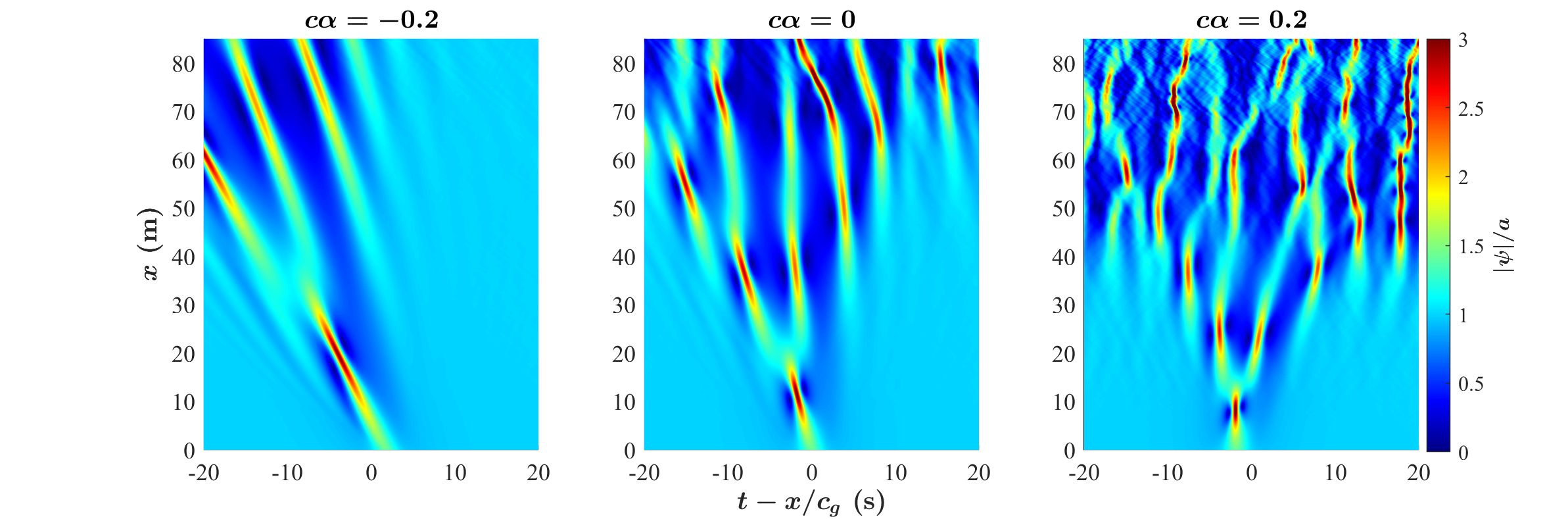}
    \includegraphics[width=1\textwidth]{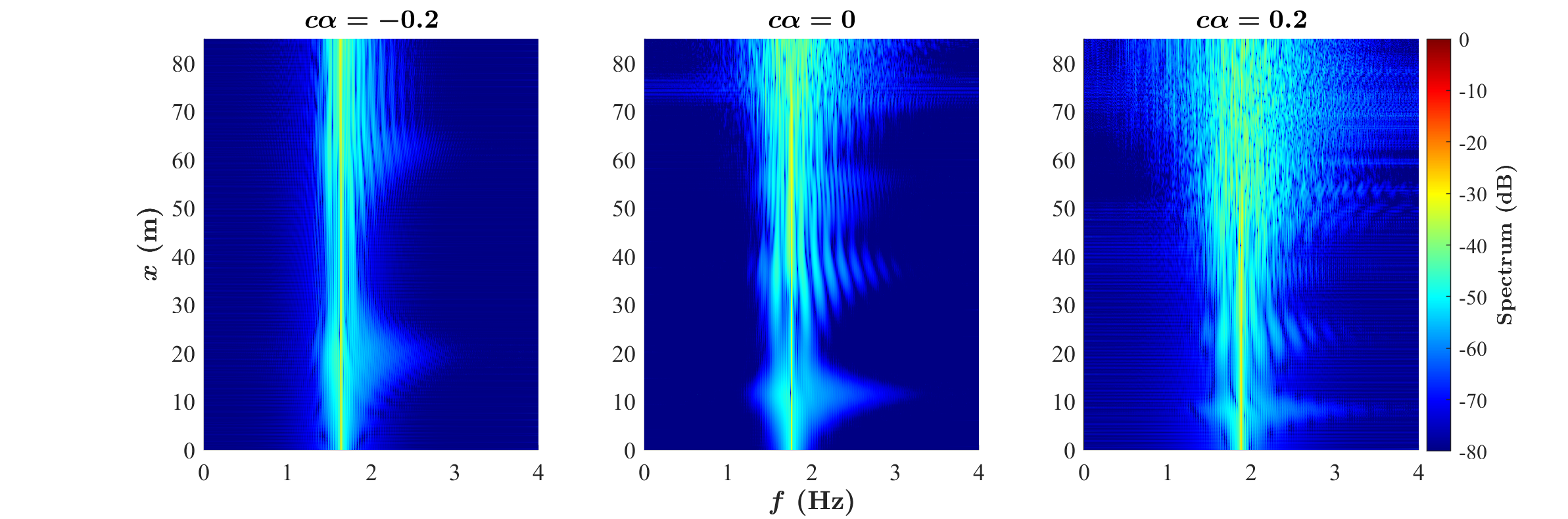}
    \caption{Long-term MNLSE prediction of Galilean-transformed Peregrine breather for the same parameters as in Fig. \ref{Fig6}. Top panels: Envelope evolution. Bottom panels: Corresponding spectral evolution.}
    \label{Fig7}
\end{figure}

The propagation of a Peregrine breathers being converted according to a GT with negative and positive GVs in the wave flume together with the respective MNLSE expectations are shown in Fig. \ref{Fig6}. 
Indeed, we can observe that the Peregrine perturbation focuses later in the case of a negative GV and earlier for the positive GV while the MNLSE simulations agree well with the observations \cite{slunyaev2013simulations,shemer2013peregrine}. However, there is an omnipresent MI dynamics, which cannot be captured within the short fetch. Hence, we extended the propagation distance of the numerical simulations to inspect the long-term behavior of the Galilean transformed Peregrine soliton following the parameters adopted in the experiments, see Fig. \ref{Fig7}. 
The MNLSE simulations suggest that long-term evolution of the Peregrine soliton is indeed complex, particularly, when GV is positive. Considering the Peregrine breather being a particular case of Kuznetsov-Ma \cite{kuznetsov1977solitons,ma_solutions_1979} and Akhmediev \cite{akhmediev1985generation} breathers, an extensive and comprehensive study is required to address and quantify the influence of the GT on solitons on finite background.

\section{Galilean-transformed envelope solitons and multi-solitons}

To disclose the relationship between multi-solitons, or Satsuma-Yajima solitons \cite{satsuma1974b}, and the Galilean-transformed solitons, we will assume that both parametrizations are equal at $x=0$, for simplicity. We recall that the dynamics of a multi-soliton $\psi_{MS}$ can be triggered by an integer-multiple of an envelope sech-type soliton \cite{dudley2013supercontinuum}. That is
\begin{equation}
    \psi_{MS}(0,t)=N\psi_s^\prime(0,t)=Na^\prime\ {\rm sech}(-\sqrt{2}\alpha^\prime k^\prime(-c_g^\prime t)),\ N\in\mathbb{N,}
\end{equation}
where $\psi_s^\prime(0,t)$ is a pure envelope soliton of amplitude $a^\prime$, wavenumber $k^\prime$ and wave frequency $\omega^\prime$. Let us assume that the GT of an envelope soliton of amplitude $a$, wave number $k$, and wave frequency $\omega$ corresponds to a real-multiple of a soliton of amplitude $a^\prime$, and a wave frequency $\omega^\prime$ which corresponds to the wave frequency of the Galilean-transformed soliton (incl. the phase-shift), while $k^\prime={w^\prime}^2/g$ 
\begin{equation}\label{eqn-Nsech=sech_galilean}
   R\psi_S'(0,t)=\psi_{GS}(0,t),\ R\in\mathbb{R}
 \end{equation}
Since the water surface elevation to first-order in steepness should be identical, the following holds
\begin{equation}\label{eqn-Nsech=sech_galilean_real}
   \eta^\prime(0,t)=\Re(R\psi_S'(0,t)\exp(i(-\omega' t)))=\Re(\psi_{GS}(0,t)\exp(i(-\omega t)))=\eta_{GS}(0,t),
 \end{equation}
which is equivalent to:
\begin{equation}
\begin{split}
Ra'\ &{\rm sech}(-\sqrt{2}\alpha' k'(-c_g't)))\exp(i(-\omega' t))=\\
 a\ &{\rm sech}(-\sqrt{2}\alpha k(-c_gt)))\exp(i(-\omega t))\exp(\frac{i\sqrt{2}c\alpha k(-c_gt)}{2}).
\end{split}
\label{eq_suface}
\end{equation}

One way to solve this is problem is to set the amplitudes, sech-components, and phases equal. Consequently, we can establish the following set of three equations 
\begin{equation}
\begin{cases} 
Ra^\prime=a\\
 \alpha' k' c_g't =\alpha k c_gt\\
 \omega' t=\omega t+\dfrac{c\alpha k c_gt}{\sqrt{2}}.
\end{cases} 
\label{eqns_solv}
\end{equation}

\begin{figure}[ht]
\centering
\includegraphics[width=1\textwidth]{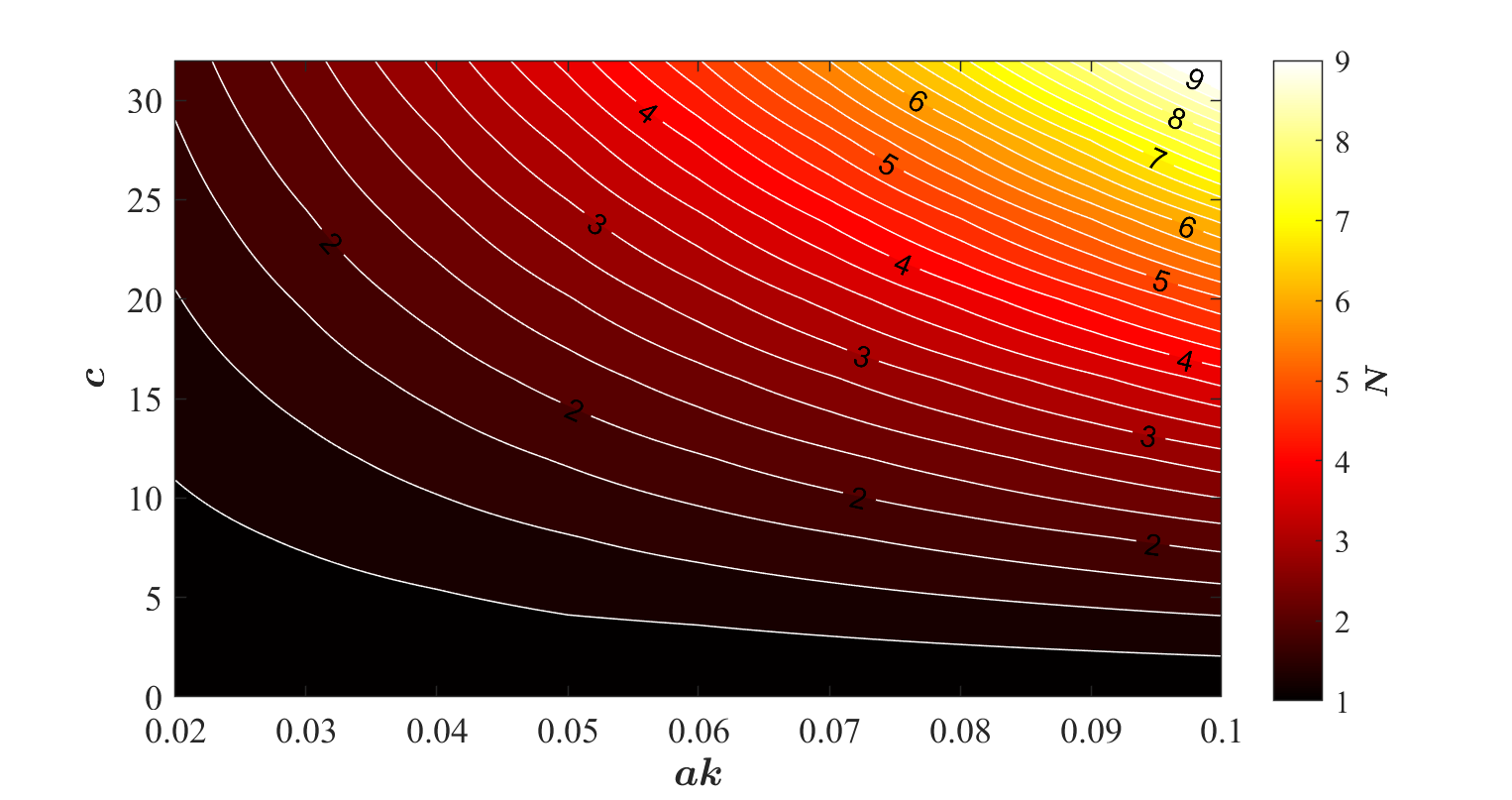}
\caption{The order factor $N$ of higher-order solitons corresponding to the Galilean solitons versus GS $c$ and steepness values $ak$.}
\label{Fig8}
\end{figure}

Using the deep-water expression for the group velocity $c_g=\dfrac{\omega}{2k}$, we can solve Eqs. (\ref{eqns_solv}) and get 
\begin{equation}\label{eqn-expression of N and omega'}
    \omega'=\omega(1+\frac{\sqrt{2}}{4}c\alpha)\ \textnormal{and}\ R=(1+\frac{\sqrt{2}}{4}c\alpha)^3.
\end{equation}
Eq. (\ref{eqn-expression of N and omega'}) shows that both, steepness and GV affect the values of $R$ and $\omega^\prime$. Moreover, these later parameters $\alpha$ and $c$ can be chosen so that $R$ becomes an integer $R=N$ and as such, Galilean-transformed solitons can define boundary conditions to launch exact multi-soliton orbits. Fig. \ref{Fig8} illustrates the range for the values of $ak$ and $c$ to map the Galilean-transformed solitons to exact Satsuma-Yajima wave groups of order $N$.   

\section{Supercontinuum generation}

Having established the relationship between Galilean transformed envelope solitons and higher-order solitons, it is self-evident to discuss supercontinuum generation following a soliton-GT. Supercontiua can be formed as a result of soliton fission of higher-order solitons due to the strong unbalance between nonlinearity and dispersion, which arises from the considerable envelope compression or focusing, i.e. substantial high values of nonlinearity \cite{dudley2006supercontinuum,dudley2010optical,dudley2013supercontinuum,chabchoub2013hydrodynamic}. 

Recurrent multi-soliton focusing and the formation of supercontinua requires a long fetch. Since these cannot be met by our state of the art facility as illustrated in Fig. 2, we will investigate and explore such dynamics using a numerical wave tank, based on the higher-order spectral model scheme, which solves the Euler equations \cite{dommermuth1987high,ducrozet2012modified}.

\begin{figure}[htp]
\centering
\includegraphics[width=1\textwidth]{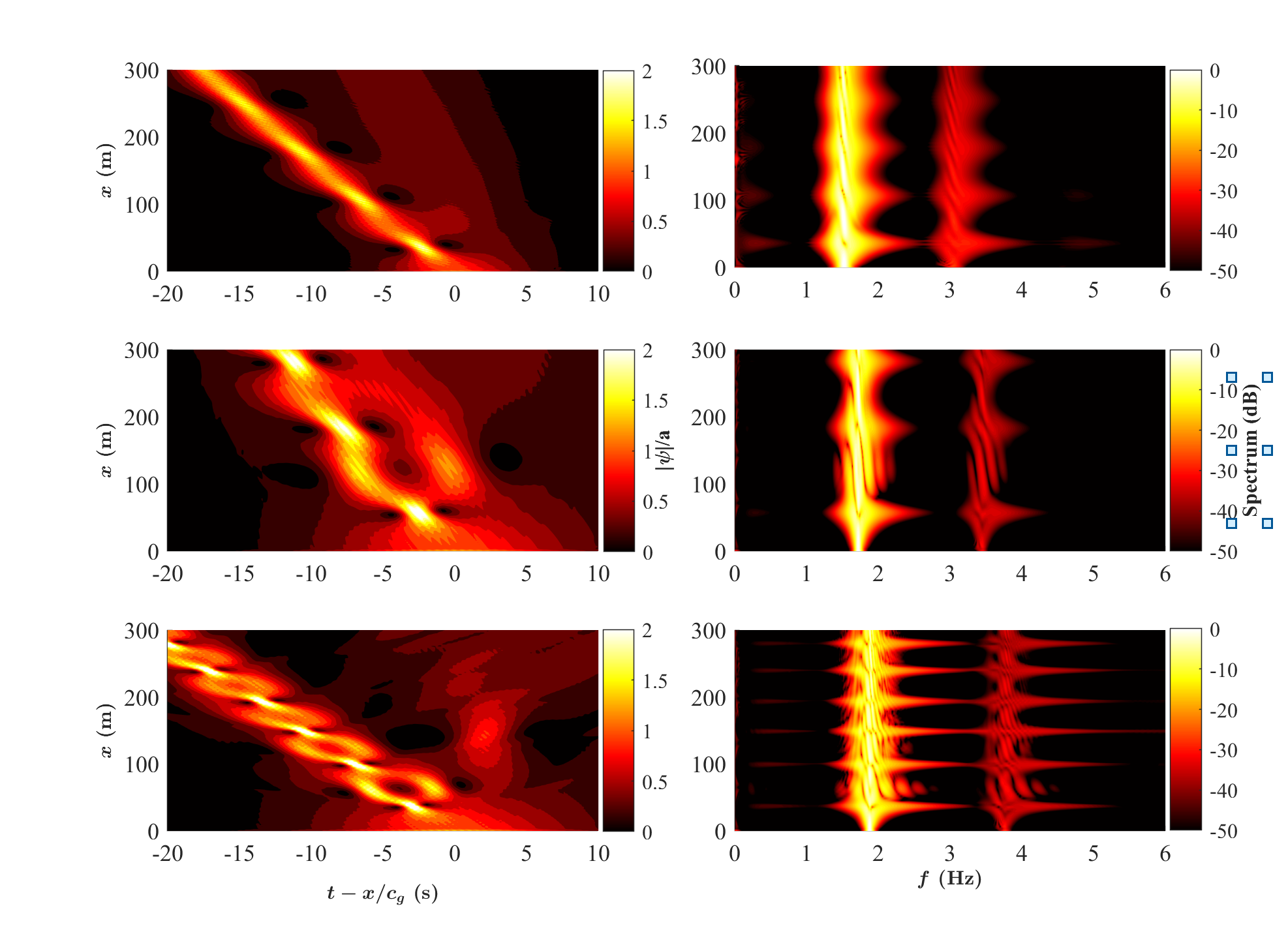}
\caption{HOSM simulations of Galilean solitons corresponding to higher-order solitons with order $N=2, 3, 4$, carrier steepness of $ak=0.075$, and frequency $f=1.25$ Hz. Top panels: $c\alpha=0.735,\ N=2$. Middle panels: $c\alpha=1.251,\ N=3$. Bottom panels: $c\alpha=1.661,\ N=4$, }
\label{Fig9}
\end{figure}

We first consider perfectly recurrent Galilean-transformed envelope solitons to be mapped on the orbit of a multi-soliton as described in the previous section. Simulation results and corresponding spectral evolution are summarized in Fig. \ref{Fig9}. 

\begin{figure}[ht]
\centering
\includegraphics[width=0.49\textwidth]{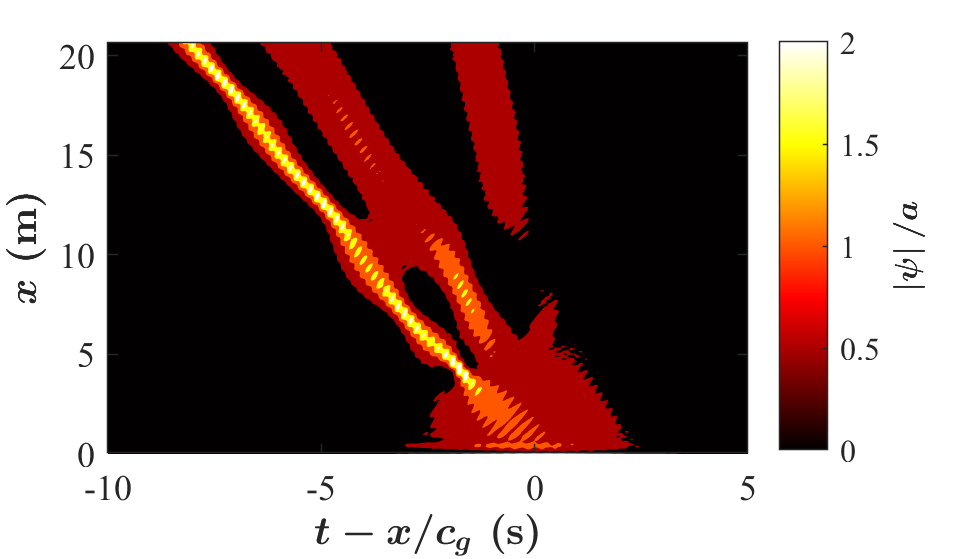}
\includegraphics[width=0.49\textwidth]{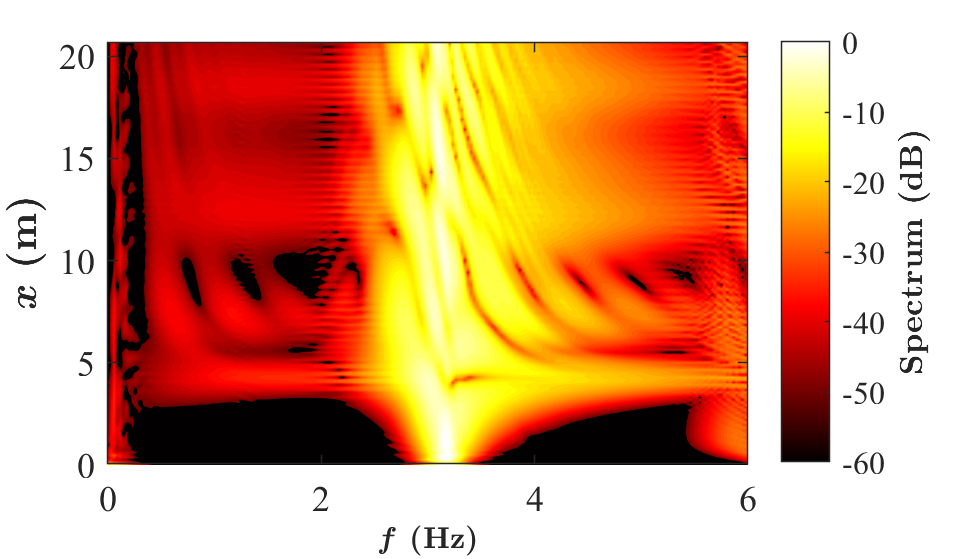}
\caption{HOSM simulations of Galilean solitons corresponding to a multi soliton of order 4 with $c\alpha=1.661$, carrier steepness $ak=0.0667$ and wave frequency $f=2.035$ Hz. The soliton fission is clearly observed.}
\label{Fig10}
\end{figure}

The HOSM simulations strikingly confirm several wave focusing recurrence cycles of wave envelope on zero-background focusing, which are clear attribute of Satsuma-Yajima breathers. This is also remarkably noticeable for the order $N=3$. One possibility to break this symmetry and induce soliton fission is to increase the order of the multi-soliton solution, i.e. the GV $c$, or the carrier wave steepness \cite{chabchoub2013hydrodynamic}. We stress that wave breaking should be excluded in this process, since the flow is assumed to be irrotational to justify the use of the HOSM. One realization involving an envelope soliton, subject to a GT and meeting an initial multi-soliton profile of the order of 4 is illustrated in Fig. \ref{Fig10}. 

The evolution of the wave envelope shows a clear soliton dominant fission signature into three solitons after the first focusing cycle. The latter mechanism brings along a severe and irreversible spectral broadening, which is a characteristic feature of a supercontiuum emergence. 



\section{Conclusion}
Galilean soliton dynamics have been investigated in controlled laboratory conditions. Even though the NLSE predicts that the GT should not affect steadiness of the envelope solitons, clear deviations from NLSE predictions have been observed and quantitatively captured by the MNLSE, which takes into account high-order dispersive effects. For positive values of GV, Galilean-transformed solitons exhibit a self-focusing dynamics, which can be in particular cases connected to exact multi-soliton solutions. Higher-order spectral method-based numerical wave tank simulations confirm such distinctive recurrent focusing dynamics of multi-solitons as obtained from GTs, including orders of 2, 3, and 4. In addition, a particular case is considered in which soliton fission has been observed involving a supercontinuum generation, as characterized by a severe and irreversible broadening of wave spectrum. Future work will be devoted to understanding and addressing the role and limitations of MNLSE in the modeling of velocity-translated solitons on zero and particularly finite background. Preliminary observations and simulation results have revealed very complex wave coherence and focusing dynamics as a result of modulation instability at play. We also expect complementary theoretical, numerical, and experimental studies in nonlinear dispersive media other than water waves. 

\section*{Acknowledgments}
J.M.D. acknowledges support from the French National Research Agency: ANR-15-IDEX-0003, ANR-17-EURE-0002, ANR-20-CE30-0004, ANR-21-ESRE-0040.

\bibliographystyle{unsrt}
\bibliography{references}
\end{document}